\newif\ifpreprint
\preprintfalse 

\ifpreprint
\documentclass[journal=jpclcd,manuscript=letter]{achemso}
\else
\documentclass[journal=jpclcd,manuscript=letter,layout=twocolumn]{achemso}
\fi

\usepackage[T1]{fontenc} 

\usepackage{amsmath}
\usepackage{newtxtext,newtxmath}

\usepackage{graphicx}
\usepackage{dcolumn}
\usepackage{braket}
\usepackage{multirow}
\usepackage{threeparttable}
\usepackage{xspace}
\usepackage{verbatim}
\usepackage[version=4]{mhchem} 
\usepackage{comment}
\usepackage{color,soul}

\usepackage[dvipsnames]{xcolor}
\usepackage{xspace}
\usepackage{ifthen}

\usepackage{qcircuit}

\usepackage{graphicx,longtable,dcolumn,mhchem}
\usepackage{rotating,color}
\usepackage{lscape}
\usepackage{amsmath}
\usepackage{dsfont}
\usepackage{soul}
\usepackage{physics}
\newcolumntype{d}{D{.}{.}{-1}}
\newcommand{\mc}{\multicolumn}

\usepackage[colorlinks = true,
            linkcolor = blue,
            urlcolor  = black,
            citecolor = blue,
            anchorcolor = black]{hyperref}

\definecolor{goodorange}{RGB}{225,125,0}
\definecolor{goodgreen}{RGB}{5,130,5}
\definecolor{goodred}{RGB}{220,50,25}
\definecolor{goodblue}{RGB}{30,144,255}

\newcommand{\note}[2]{
\ifthenelse{\equal{#1}{F}}{
\colorbox{goodorange}{\textcolor{white}{\footnotesize \fontfamily{phv}\selectfont #1}}
    \textcolor{goodorange}{{\footnotesize \fontfamily{phv}\selectfont #2}}\xspace
}{}
\ifthenelse{\equal{#1}{R}}{
\colorbox{goodred}{\textcolor{white}{\footnotesize \fontfamily{phv}\selectfont #1}}
    \textcolor{goodred}{{\footnotesize \fontfamily{phv}\selectfont #2}}\xspace
}{}
\ifthenelse{\equal{#1}{N}}{
\colorbox{goodgreen}{\textcolor{white}{\footnotesize \fontfamily{phv}\selectfont #1}}
    \textcolor{goodgreen}{{\footnotesize \fontfamily{phv}\selectfont #2}}\xspace
}{}
\ifthenelse{\equal{#1}{M}}{
\colorbox{goodblue}{\textcolor{white}{\footnotesize \fontfamily{phv}\selectfont #1}}
    \textcolor{goodblue}{{\footnotesize \fontfamily{phv}\selectfont #2}}\xspace
}{}
}

\usepackage{titlesec}

\usepackage[fontsize=11pt]{scrextend}
\titleformat{\section}
{\normalfont\sffamily\bfseries\color{Blue}}
{\thesection.}{0.25em}{\uppercase}

\titleformat{\subsection}
{\normalfont\sffamily\bfseries}
{\thesubsection}{0.25em}{}

\titleformat{\subsubsection}
{\normalfont\sffamily}
{\thesubsubsection}{0.25em}{}

\titleformat{\suppinfo}
{\normalfont\sffamily\bfseries}
{\thesubsection}{0.25em}{}

\titlespacing*{\section}{0pt}{0.5\baselineskip}{0.01\baselineskip}
\titlespacing*{\subsection}{0pt}{0.125\baselineskip}{0.01\baselineskip}
\titlespacing*{\subsubsection}{0pt}{0.125\baselineskip}{0.01\baselineskip}

\author{Pierre-Fran{\c c}ois Loos}
	\affiliation[LCPQ, Toulouse]{Laboratoire de Chimie et Physique Quantiques, Universit\'e de Toulouse, CNRS, UPS, France}
\author{Filippo Lipparini}
	\affiliation[UP, Pisa]{Dipartimento di Chimica e Chimica Industriale, University of Pisa, Via Moruzzi 3, 56124 Pisa, Italy}
\author{Denis Jacquemin}
	\email{Denis.Jacquemin@univ-nantes.fr }
	\affiliation[CEISAM, Nantes]{Nantes Universit\'e, CNRS, CEISAM UMR 6230, F-44000 Nantes, France}
	\alsoaffiliation[IUF, Paris]{Institut Universitaire de France, 75005 Paris, France}

\let\oldmaketitle\maketitle
\let\maketitle\relax
	\title{Heptazine, Cyclazine, and Related Compounds: Chemically-Accurate Estimates of the Inverted Singlet-Triplet Gap} 
\date{\today}

\begin{tocentry}
	\centering
	\includegraphics[width=1.0\textwidth]{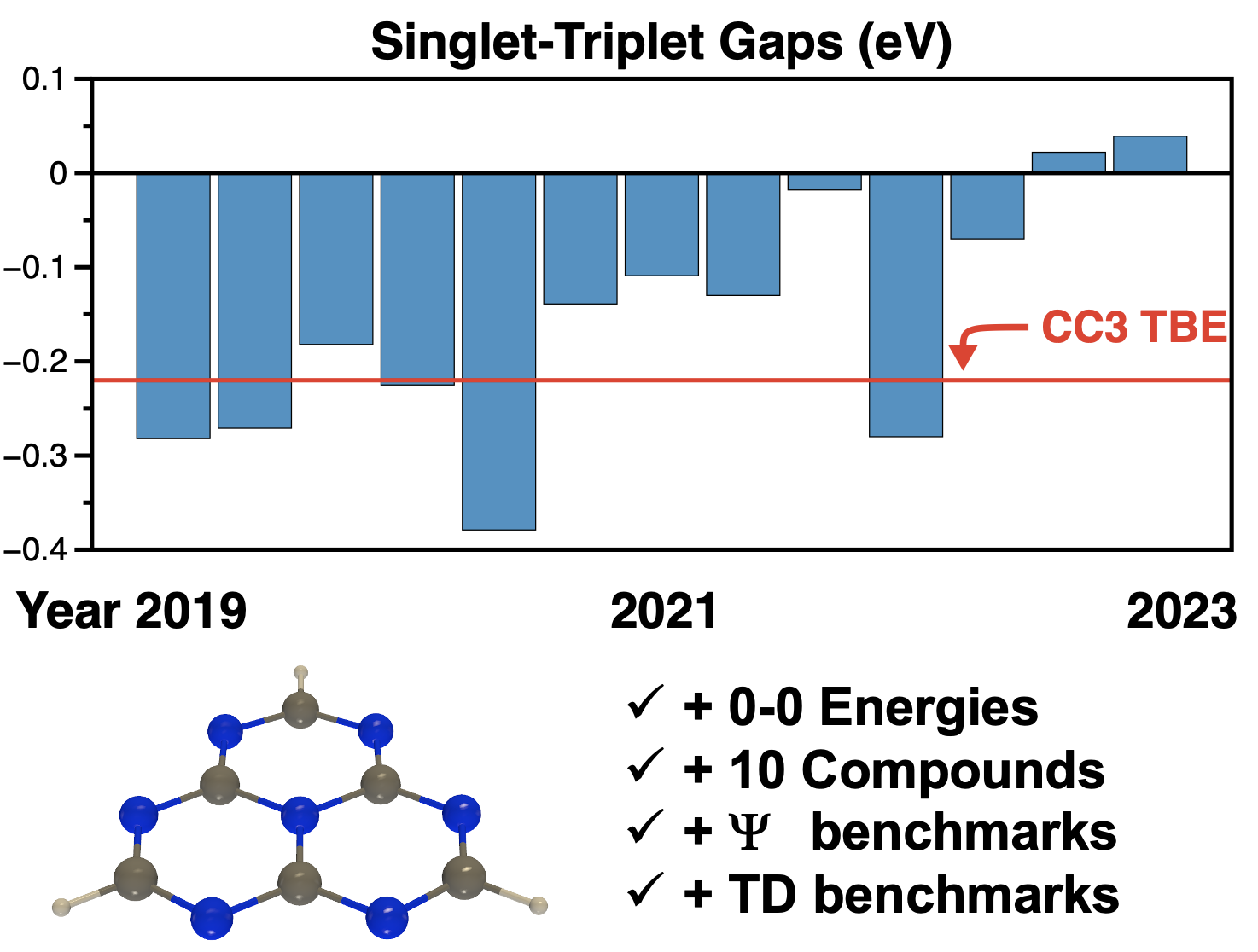}
\end{tocentry}

\begin{document}	

\ifpreprint
\else
\twocolumn[
\begin{@twocolumnfalse}
\fi
\oldmaketitle


\begin{abstract}
Molecules that violate Hund's rule and exhibit an inverted gap between the lowest singlet $S_1$ and triplet $T_1$ excited states have attracted considerable attention due to their potential applications in optoelectronics.
Amongst these molecules, the triangular-shaped heptazine, and its derivatives, have been in the limelight.
However, conflicting reports have arisen regarding the relative energies of $S_1$ and $T_1$. 
Here, we employ highly accurate levels of theory, such as CC3, to not only resolve the debate concerning the sign but also quantify the magnitude of the $S_1$-$T_1$ gap. 
We also determine the 0-0 energies to evaluate the significance of the vertical approximation. 
In addition, we compute reference $S_1$-$T_1$ gaps for a series of 10 related molecules. 
This enables us to benchmark lower-order methods for future applications on larger systems within the same family of compounds. 
This contribution can serve as a reliable foundation for the design of triangular-shaped molecules with enhanced photophysical properties.
\end{abstract}

\ifpreprint
\else
\end{@twocolumnfalse}
]
\fi

\ifpreprint
\else
\small
\fi

\clearpage

\noindent

In many technological applications that rely on either light-energy or energy-light conversions, it is essential to control the energy gap that separates the relevant electronic excited states (ESs). 
Specifically, the magnitude of the singlet-triplet gap (STG), the energy difference between the lowest-energy singlet excited state, $S_1$, and the lowest-energy triplet state, $T_1$, is of prime importance as key photophysical events typically take place 
in these states.  For example, to design highly fluorescent molecules, a large STG is sought after to minimize the $S_1 \rightarrow T_1$ intersystem crossing. 
In contrast, in the pursuit of efficient third-generation organic light-emitting diodes (OLEDs), a vanishingly small STG is required to maximize the reverse intersystem crossing ($T_1 \rightarrow S_1$), resulting in the production of more photons. 

Despite constant developments, experimental characterizations of ESs remain both costly and challenging. This is why theoretical models, which explore the ES nature and energies are frequently employed hand-in-hand 
with experimental measurements. Nevertheless, theoretical methods also face challenges in achieving chemically accurate (i.e., error of the order of $\pm 0.05$ eV) estimates of ES energies for non-trivial molecular systems. 
This is in stark contrast with ground-state (GS) properties for which a panel of reliable black-box methods is now readily available. The search for molecular structures having specific, original, and tuneable 
ES properties are therefore extremely active research lines. 

\begin{figure}[ht]
\centering
	 \includegraphics[width=.55\linewidth]{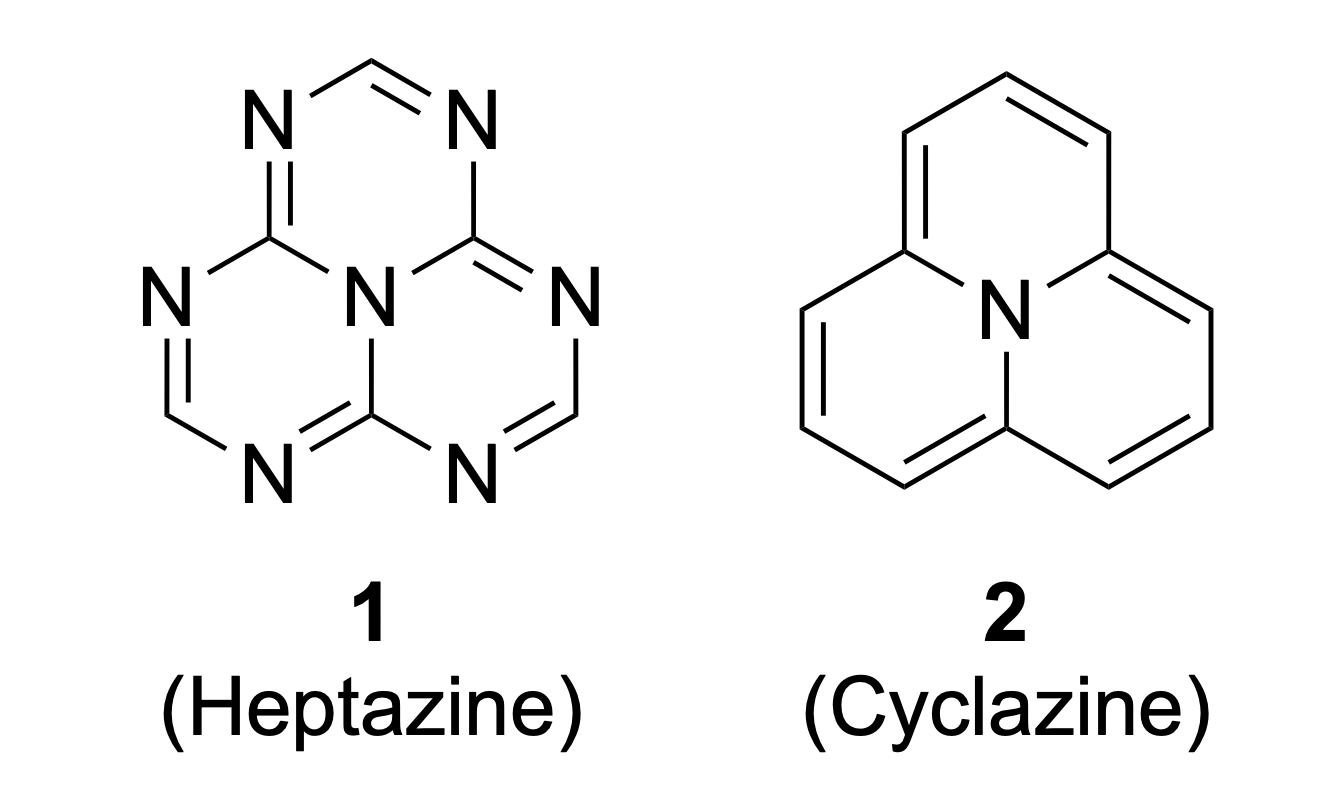}
	\caption{Representation of heptazine and cyclazine.}
	\label{Fig-1}
\end{figure}

In 2019, two groups independently published seminal works showing that some triangulenes do not adhere to Hund's rule, with $T_1$ being found above $S_1$. \cite{Ehr19,Des19}. On the one hand,
Ehrmaier and coworkers demonstrated that this inversion takes place in heptazine (compound {\bfseries 1} in Figure \ref{Fig-1}), \cite{Ehr19} whereas, on the other hand, de Silva reported negative STG in
cyclazine (compound {\bfseries 2} in Figure \ref{Fig-1}). \cite{Des19} These  remarkable and unexpected discoveries, supported by both theoretical and experimental analyses (as discussed below), \cite{Ehr19,Des19} 
stimulated many subsequent works. \cite{Ric21,Pol21,Din21,Gho22,Aiz22,Ric22,Tuc22,Mon23,Drw23,Bla23,Dre23} In particular, several groups analyzed in detail the reasons behind this deviation from Hund's rule,
 \cite{Ric21,Ric22,Bla23,Drw23,San23} and concurrently, improved substituted molecules have been designed using various modeling approaches. \cite{Pol21,San21,Ric22,Sob23}

In the original work of Ehrmaier \emph{et al.}, a series of wave-function calculations performed with the double-$\zeta$ basis set cc-pVDZ was used to estimate the STG of heptazine 
(see Table \ref{Table-1}). 
This investigation revealed that time-dependent density-functional theory (TD-DFT) relying on global hybrids --- the typical workhorse for ES calculations --- failed to deliver 
qualitatively correct values. \cite{Ehr19} 
Notably, the most sophisticated levels of theory used in Ref.~\citenum{Ehr19}, namely, coupled-cluster with singles and doubles (CCSD) and the complete-active-space self-consistent-field method
with second-order perturbative correction (CASPT2), provided smaller STG than less advanced second-order approaches (Table \ref{Table-1}). 

\begin{table}[htp]
\caption{\footnotesize Vertical transition energies to the lowest $S_1$ and $T_1$ ESs, as well as the corresponding STG of heptazine obtained with various levels of theory. 
All values are in eV.$^a$}
\label{Table-1}
\footnotesize
\vspace{-0.2 cm}
\begin{tabular}{lccccc}
\hline
Method						& $S_1$	& $T_1$	& STG	& Ref.	&	Year\\
\hline						
ADC(2)/cc-pVDZ				&2.569	&2.851	&-0.282	& \citenum{Ehr19} & 2019\\
CC2/cc-pVDZ					&2.676	&2.947	&-0.271	& \citenum{Ehr19} & 2019\\
CCSD/cc-pVDZ					&2.791	&2.963	&-0.182	& \citenum{Ehr19} & 2019\\
CASPT2/cc-pVDZ				&2.326	&2.551	&-0.225	& \citenum{Ehr19} & 2019\\
SCS-CC2/def2-TZVP			&2.847	&3.226	&-0.379	& \citenum{Ric21} & 2021\\
SC-NEVPT2/def2-TZVP			&3.259	&3.398	&-0.139	& \citenum{Ric21} & 2021\\
ADC(3)/cc-pVDZ				&2.665	&2.774	&-0.109	& \citenum{Pol21} & 2021 \\
RASPT2/def2-TZVP				&2.54	&2.67	&-0.13	& \citenum{Gho22} & 2022 \\
XMC-QDPT2/def2-TZVP			&		&		&-0.018	& \citenum{Tuc22} & 2022 \\
Mk-MRCCSD(T)/def2-TZVP		&		&		&-0.28	& \citenum{Drw23} & 2023\\
ADC(3)/cc-pVTZ				&2.81	&2.88	&-0.07	& \citenum{Dre23} & 2023\\
$\Delta$CCSD(T)/\emph{aug}-cc-pVDZ&		&		&+0.022	& \citenum{Dre23} & 2023\\
$\Delta$CCSD(T)/cc-pVTZ		&		&		&+0.039	& \citenum{Dre23} & 2023\\
\hline
ADC(2)/\emph{aug}-cc-pVTZ		&2.675	&2.921	&-0.246	&	\mc{2}{c}{This work}\\
CC2/\emph{aug}-cc-pVTZ			&2.767	&3.006	&-0.239	& \mc{2}{c}{This work}\\
CCSD/\emph{aug}-cc-pVTZ		&2.953	&3.087	&-0.134	& \mc{2}{c}{This work}\\
CC3/\emph{aug}-cc-pVDZ			&2.693	&2.898	&-0.205	& \mc{2}{c}{This work}\\
TBE$^b$						&2.717	&2.936	&-0.219	& \mc{2}{c}{This work}\\
\hline
\end{tabular}
\begin{scriptsize}
\vspace{-0.2 cm}
\begin{flushleft}
$^a${The present calculations are performed on the CCSD(T)/cc-pVTZ GS structure. Note that previous works used (slightly) different geometries.}
$^b${Theoretical best estimate obtained from CC3/\emph{aug}-cc-pVTZ + [CCSDT/6-31+G(d) $-$ CC3/6-31+G(d)] for $S_1$ and CC3/\emph{aug}-cc-pVDZ + [CCSD/\emph{aug}-cc-pVTZ $-$ CCSD/\emph{aug}-cc-pVDZ] for $T_1$.}
\end{flushleft}
\end{scriptsize}
\end{table}

Two years later, Ricci \emph{et al.} relied on a significantly larger 
basis set, def2-TZVP,  and reported a STG of $-0.139$ eV at the $N$-electron valence state second-order perturbation theory (NEVPT2) level using a large $(12,12)$ active space. \cite{Ric21}  The STG value of Ghosh and Bhattacharyya computed with the restricted-active-space method (RASPT2) is similar to the NEVPT2 value ($-0.13$ eV), whereas similarity-transformed equation-of-motion CCSD (STEOM-CCSD) unexpectedly yields a much 
larger result ($-0.66$ eV). \cite{Gho22}  A comparison between the multireference values of Refs.~\citenum{Ehr19}, \citenum{Ric21}, and \citenum{Tuc22} also hints that increasing the size of the basis set from 
cc-pVDZ to def2-TZVP induces a significant drop of the STG magnitude.  
In 2022, new experiments were carried out for two substituted heptazines, and a tiny yet negative gap ($-0.011$ eV) was found for one of them, \cite{Aiz22}  whereas very recently, Actis \textit{et al.}~reported a negative STG of approximately $-0.2$ eV in an extended carbon-nitride (CN) graphite-like structure. \cite{Act23}

Very recently, Dreuw and Hoffmann went a step further by questioning the existence of an inverted STG in heptazine and two closely related molecules (cyclazine, {\bfseries{2}}, and cyclborane). \cite{Dre23} 
By systematically ramping up the degree of the algebraic-diagrammatic construction (ADC), going from ADC(2)-s, to ADC(2)-x, and then ADC(3), they observed clear reductions of the STGs, 
with an ADC(3)/cc-pVTZ gap amounting to a mere $-0.07$ eV. In addition, they conducted (state-specific) $\Delta$CC calculations based on non-Aufbau reference determinants with a series of increasingly larger basis sets and obtained a positive STG of $+0.04$ eV with 
their most refined approach, namely, $\Delta$CCSD(T)/cc-pVTZ.  
In short, as the level of theory became more accurate, the negative character of the STG in heptazine diminished.  
Dreuw and Hoffmann, therefore, concluded that the sign (amplitude) of the STG of heptazine was uncertain  (notably small), emphasizing the necessity for more accurate calculations \cite{Dre23} This Letter answers their call.

To obtain very accurate estimates of the ES energies of heptazine, i.e., theoretical best estimates (TBEs), we employed CC methodologies including singles, doubles, and iterative triples, along with large basis sets (see computational details section). 
This approach represents a substantial advancement in comparison to earlier studies. 
First, we optimized the $D_{3h}$ GS structure of heptazine at the CCSD(T)/cc-pVTZ level, ensuring a solid starting point for further calculations. Next, we performed a series 
of equation-of-motion CC (EOM-CC) calculations with various diffuse-containing basis sets, up to third-order CC (CC3) for the triplet state and even using full triples (CCSDT) for the singlet state (see the SI for the complete set of
data). Importantly, we did not consider state-specific approaches, such as $\Delta$SCF and $\Delta$CC, but systematically relied on linear response theory to directly target ES energies. 

It is important to note that the so-called GS $T_1$-diagnostic \cite{Lee89} returns a value of 0.02 at the CCSD/\emph{aug}-cc-pVTZ level for heptazine, indicating the absence of significant multireference character in the GS. 
Furthermore, our CC3/\emph{aug}-cc-pVDZ reveals no substantial double excitation character with $\%T_1$ values (which indicates the percentage of single excitation character) of 86.3\% and 95.7\% for the lowest excited singlet and triplet states, respectively.  
These observations are consistent with both previous conclusions drawn at the second-order CC level, \cite{Tuc22} and the negligible difference observed between CC3 and CCSDT for the $S_1$ excitation energy ($< 0.01$ eV, see the SI). 
There is therefore no clear reason to delve into multiconfigurational methodologies here, as the CC family of systematically improvable methods appears to offer the most efficient path to chemical accuracy. 

In the extended \textsc{quest} database, \cite{Ver21} it was found that the typical error associated with CC3 is as small as $0.02$ eV for such well-behaved ES (having $\%T_1 > 85\%$). 
We are therefore reasonably confident that the data reported in Table \ref{Table-1} stand as the most reliable to date and are likely chemically accurate (error below $0.05$ eV). 
As reported in Table \ref{Table-1}, our TBE for the singlet and triplet states are respectively $2.717$ and $2.936$ eV, resulting in a negative gap of $-0.219$ eV. It is no surprise that this value is in between the CC2 and CCSD results obtained with large basis sets. 
Second-order methods such as ADC(2) and CC2 tend to overestimate the magnitude of the STG, while CCSD and ADC(3) exhibit a tendency to provide an error in the opposite direction.
We note that in the  \textsc{quest} database, \cite{Ver21} CCSD was found to overshoot the triplet (singlet) energies by $0.03$ ($0.14$) eV on average, whereas ADC(3) was found to underestimate triplet (singlet) excitation energies by $-0.18$  ($-0.08$) eV,
which is consistent with the present trend.
We also note that previous CASPT2 (NEVPT2) significantly underestimate (overestimate) the absolute energies of $S_1$ and $T_1$. Finally, it appears that the \emph{aug}-cc-pVDZ basis set proves sufficiently large to deliver accurate results (at the CC3 level) while the cc-pVDZ basis set produces an excessively negative STG. 
In any case, the data showcased in Table \ref{Table-1} conclusively affirm that the vertical excitation energy of $T_1$ in heptazine is approximately $0.22$ eV higher than that of $S_1$.

In Table \ref{Table-2}, we provide a similar comparative analysis, juxtaposing our present estimates with prior data from the literature for the singlet and triplet vertical excitation energies of cyclazine (compound {\bfseries 2} in Figure \ref{Fig-1}). 
While the two states are notably closer to the GS than in heptazine, the methodological trends outlined in the previous paragraph for heptazine mostly pertain. For example, using a small basis set yields too negative STGs, whereas 
CCSD and ADC(3) underestimate the gap. For cyclazine, both ADC(2) and CC2 overestimate the excitation energies, but they provide an excellent estimate of the STG, SCS-CC2 being significantly off.
Our TBE for the STG in cyclazine is $-0.131$ eV, which is large enough to affirm the gap inversion in this system.
 
\begin{table}[htp]
\caption{\footnotesize Vertical transition energies to the lowest $S_1$ and $T_1$ ESs, as well as corresponding STG of cyclazine obtained with various levels of theory. All values are in eV.$^a$}
\vspace{-0.2 cm}
\label{Table-2}
\footnotesize
\begin{tabular}{lccccc}
\hline
Method						& $S_1$	& $T_1$	& STG	& Ref. & Year\\
\hline		
CIS(D)/cc-pVDZ				&1.07	&1.37	&-0.30	&\citenum{Des19} & 2019\\				
ADC(2)/cc-pVDZ				&1.04	&1.20	&-0.16	&\citenum{Des19} & 2019\\
CCSD/cc-pVDZ					&1.09	&1.19	&-0.10	&\citenum{Des19} & 2019\\
SCS-CC2/def2-TZVP			&1.110	&1.334	&-0.224	&\citenum{Ric21} & 2021\\
SC-NEVPT2/def2-TZVP			&1.224	&1.288	&-0.044	&\citenum{Ric21} & 2021\\
ADC(3)/cc-pVDZ				&0.777	&0.869	&-0.092	&\citenum{Pol21} & 2021 \\
RASPT2/def2-TZVP				&0.86	&0.89	&-0.03	&\citenum{Gho22} & 2022 \\
XMC-QDPT2/def2-TZVP			&		&		&-0.106	&\citenum{Tuc22} & 2022 \\
CC3/cc-pVDZ					&0.98	&1.15	&-0.17	&\citenum{Mon23} & 2023\\
Mk-MRCCSD(T)/def2-TZVP		&		&		&-0.18	&\citenum{Drw23} & 2023 \\
ADC(3)/cc-pVTZ				&0.81	&0.87	&-0.06	&\citenum{Dre23} & 2023\\
$\Delta$CCSD(T)/\emph{aug}-cc-pVDZ&		&		&+0.015	&\citenum{Dre23} & 2023\\
$\Delta$CCSD(T)/cc-pVTZ		&		&		&+0.025	&\citenum{Dre23} & 2023\\
\hline
ADC(2)/\emph{aug}-cc-pVTZ		&1.001	&1.138	&-0.137	&\mc{2}{c}{This work}\\
CC2/\emph{aug}-cc-pVTZ			&1.051	&1.181	&-0.130	&\mc{2}{c}{This work}\\
CCSD/\emph{aug}-cc-pVTZ		&1.090	&1.154	&-0.064	&\mc{2}{c}{This work}\\
CC3/\emph{aug}-cc-pVDZ			&0.990	&1.121	&-0.131	&\mc{2}{c}{This work}\\
TBE$^b$						&0.979	&1.110	&-0.131	&\mc{2}{c}{This work}\\
\hline
\end{tabular}
\vspace{-0.2 cm}
\begin{scriptsize}
\begin{flushleft}
$^a${The present calculations are performed on the CCSD(T)/cc-pVTZ GS structure. Note that previous works used (slightly) different geometries. The experimental absorption of cyclazine in hexane was measured at $0.972$ eV. \cite{Leu80}}
$^b${Theoretical best estimate obtained from CC3/\emph{aug}-cc-pVTZ + [CCSDT/6-31+G(d) $-$ CC3/6-31+G(d)] for $S_1$ and CC3/\emph{aug}-cc-pVDZ + [CCSD/\emph{aug}-cc-pVTZ $-$ CCSD/\emph{aug}-cc-pVDZ] for $T_1$.}
\end{flushleft}
\end{scriptsize}
\end{table}

Let us now turn towards the 0-0 energies of heptazine, as this property provides a reliable basis for comparisons with experimental data. \cite{Loo19b} For obvious computational reasons, we use CCSD/cc-pVDZ, as well as its EOM-CCSD
and unrestricted variants (UCCSD) to determine the geometries and vibrational frequencies of $S_0$, $S_1$, and $T_1$, respectively. 
While the geometries obtained are not as accurate as those computed at the CCSD(T)/cc-pVTZ level (see above and below), it should be stressed that the 0-0 energies are known to be less sensitive to the accuracy of the employed structures than vertical excitation energies. \cite{Loo19a} 
Besides, the CC3/\emph{aug}-cc-pVDZ vertical  $S_1$ excitation energy computed on the CCSD/cc-pVDZ geometry is $2.649$ eV, which is quite close to the $2.693$ eV result obtained with the CCSD(T)/cc-pVTZ structure. 
Nevertheless, the use of different levels of theory for the singlet and triplet geometries likely results in a 0-0 STG value significantly less accurate than its vertical counterpart.

\begin{figure}[ht]
\centering
	 \includegraphics[width=1.0\linewidth,viewport=2cm 21.5cm 19cm 27cm,clip]{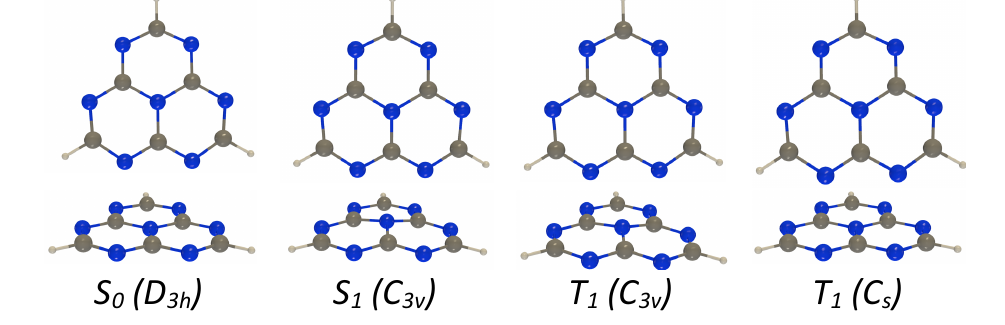}
	\caption{Two different views of the CCSD, EOM-CCSD, and UCCSD optimized geometries obtained with the cc-pVDZ basis set.}
	\label{Fig-2}
\end{figure}

Starting from the planar $D_{3h}$ GS geometry, the EOM-CCSD optimization of $S_1$ led to one imaginary frequency associated with the puckering of the central nitrogen atom. The true minimum of the $S_1$ structure presents a $C_{3v}$ 
point group symmetry (see Figure \ref{Fig-2}). The CC3/\emph{aug}-cc-pVTZ  $S_0$-$S_1$  adiabatic energy, $E^{\text{adia}}$, is $2.550$ eV, which can be corrected by the difference of zero-point vibrational 
energies between both states ($\Delta E^{\text{ZPVE}}=-0.044$ eV) determined at the (EOM-)CCSD/cc-pVDZ level to obtain a 0-0 TBE of $2.506$ eV for the  $S_0-S_1$ transition.
The difference of approximately $-0.2$ eV between the 0-0 and vertical transition energies  aligns with the typical corrections observed for relatively rigid molecules. \cite{Goe09,Jac12d,Win13}  
On a methodological note, at the CCSD(T)(a)$^\star$/\emph{aug}-cc-pVTZ level, we get $E^{\text{adia}} = 2.671$ eV ($E^{\text{0-0}}=2.627$ eV), for the $S_0$-$S_1$ transition,  a value $0.12$ eV larger than the CC3 result. 
This data is likely more directly comparable to the results obtained for the triplet state below, for which CCSD with perturbative triples is used to obtain the best estimate.

For the lowest triplet,  the $D_{3h}$ optimal structure shows two imaginary frequencies at the UCCSD/cc-pVDZ level. 
The first one leads to a puckered $C_{3v}$ geometry similar to the $S_1$ case, and the second one corresponds to an in-plane deformation providing a $C_s$ geometry, the $C_{3h}$ structure being unstable (Figure \ref{Fig-2}).  
Both the $C_{3v}$ and $C_s$ structures are genuine minima and have quite similar energies. For the former, the UCCSD(T)/\emph{aug}-cc-pVTZ value for $E^{\text{adia}}$ is $3.154$ eV and the CCSD-based ZPVE correction is quite large ($-0.189$ eV) leading to a 0-0 transition at $2.965$ eV. 
For the latter, we get $E^{\text{adia}} = 3.038$ eV and $E^{\text{0-0}} = 3.057$ eV, this structure showing an uncommon positive ZPVE correction. 
In any case, the 0-0 energies of these two triplet structures are clearly larger than the $2.627$ eV estimate obtained with a comparable level of theory for $S_1$. 
In other words, the negative STG obtained from the vertical transitions ($-0.219$ eV) does not change sign when considering geometry relaxation and zero-point vibrational effects. 
It apparently seems to become slightly larger ($-0.338$ eV), though we acknowledge that the latter value is likely not chemically accurate due to the increased number of approximations made.
Nevertheless, previous works also observed that the inverted STG found through vertical calculations persists when considering 0-0 energies. \cite{Ehr19,Pol21,Sob21}

Of course, the state-of-the-art methods discussed above are not readily applicable to the design of novel compounds with enhanced properties, a goal typically achieved through the addition of side substituents or the extension of the conjugated core. 
\cite{Pol21,Pio21,Sob21,San21,Ric22,Sob23} A wide panel of strategies has been explored in the literature to design new compounds, but the challenge lies in the absence of definitive benchmark values. This lack of a clear reference is evident, for instance, 
in the varying choices of reference methods made by different research groups when benchmarking double-hybrid functionals, with some using  SCS-CC2\cite{San22} and others opting for CCSD.\cite{Ali22,Cur23} 

\begin{figure}[ht]
\centering
	 \includegraphics[width=1.0\linewidth]{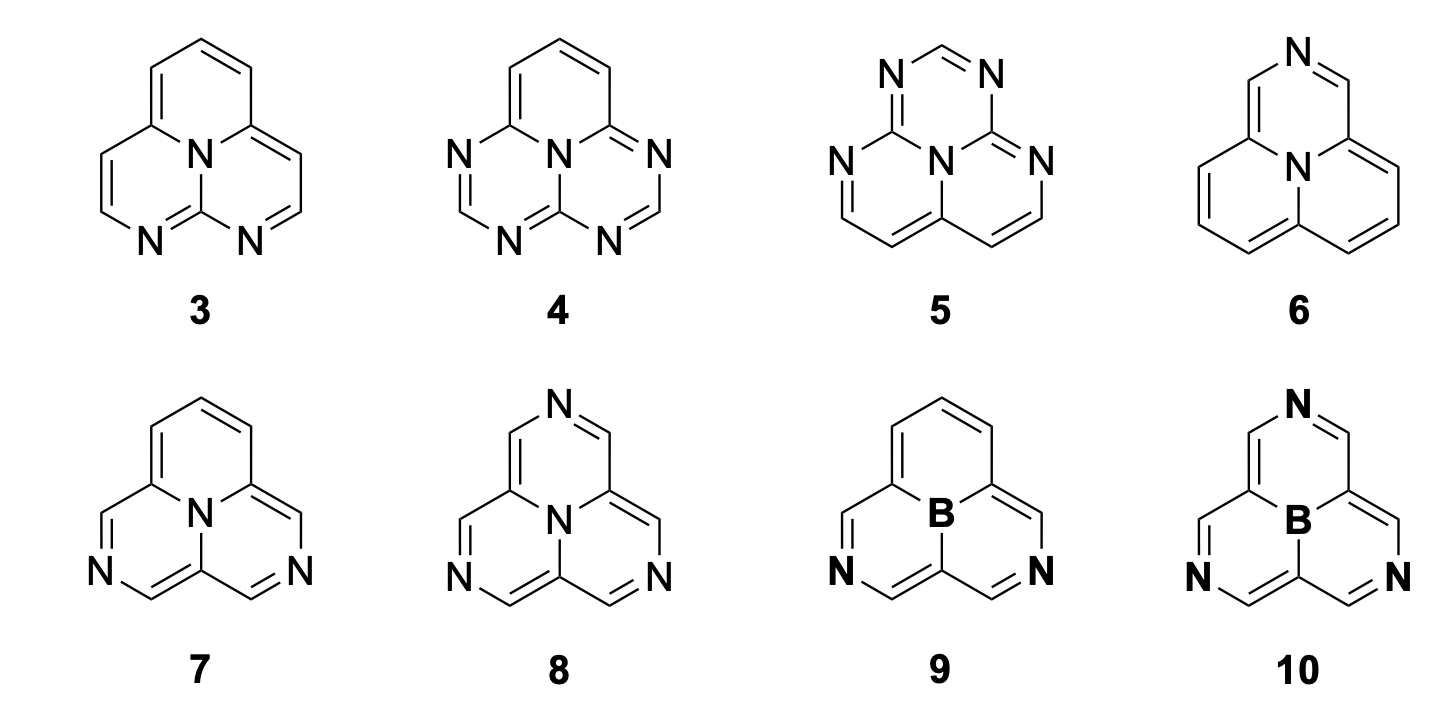}
	\caption{Representation of the eight additional molecules considered in the present benchmark set.}
	\label{Fig-3}
\end{figure}

Hence, we decided to perform a benchmark of lower-order approaches for a set of 10 triangulenes, including heptazine, cyclazine, and eight additional molecules displayed in Figure \ref{Fig-3}, for which we obtained accurate 
TBE/\emph{aug}-cc-pVTZ values using the same procedure as in Tables \ref{Table-1} and \ref{Table-2}.  The selection of these molecules was inspired by previous literature. \cite{San22,Tuc22,Drw23} We underline that the magnitude 
of the STG varies significantly within this series, as do the absolute singlet and triplet excitation energies (see Table S1 in the SI). The TBEs for the STG, listed in Table \ref{Table-3}, range from $-0.029$ eV ({\bfseries{8}}) to 
$-0.305$ eV ({\bfseries{10}}) and thus cover a significant range of possibilities. With these TBE values in hand, we conducted an evaluation of the performance of second-order wave function methods, specifically, CIS(D), ADC(2), CC2, 
and some of their spin-scaled variants (SOS-ADC(2), SCS-ADC(2), SOS-CC2 and SCS-CC2), as well as CCSD. We also benchmarked double hybrid functionals within the TD-DFT framework. More specifically, we tested four of 
them, namely PBE0-2, SOS-PBE-QIDH, SCS-PBE-QIDH,  and SOS-RSX-QIDH, as these were previously identified as promising candidates for this particular task in a benchmark study performed by Sancho-Garcia, Adamo, and coworkers. 
\cite{San22} All results can be found in Tables S2 and S3 of the SI.  

\begin{table}[htp]
\caption{\footnotesize Theoretical best estimates obtained for all compounds considered here. All values are in eV.$^a$}
\vspace{-0.2 cm}
\label{Table-3}
\footnotesize
\begin{tabular}{p{2cm}p{2cm}p{2cm}p{2cm}}
\hline
Compound			& $S_1$	& $T_1$	& STG	\\
\hline		
{\bfseries 1}		&$2.717$&$2.936$	&$-0.219$	\\
{\bfseries 2}		&$0.979$&$1.110$	&$-0.131$	\\
{\bfseries 3}		&$1.562$&$1.663$	&$-0.101$	\\
{\bfseries 4}		&$2.177$&$2.296$	&$-0.119$	\\
{\bfseries 5}		&$2.127$&$2.230$	&$-0.103$	\\
{\bfseries 6}		&$0.833$&$0.904$	&$-0.071$	\\
{\bfseries 7}		&$0.693$&$0.735$	&$-0.042$	\\
{\bfseries 8}		&$0.554$&$0.583$	&$-0.029$	 \\
{\bfseries 9}		&$1.264$&$1.463$	&$-0.199$	\\
{\bfseries 10}		&$1.522$&$1.827$	&$-0.305$	\\
\hline
\end{tabular}
\vspace{-0.2 cm}
\begin{scriptsize}
\begin{flushleft}
$^a${See Table S1 in the SI for details and raw data.}
\end{flushleft}
\end{scriptsize}
\end{table}

The results of this benchmark for the STG are displayed in Table \ref{Table-4} in which we report the mean signed error (MSE), the mean absolute error (MAE), and the standard deviation of the errors (SDE) using our TBEs as reference.
Equivalent statistical analyses for the singlet and triplet energies themselves can be found in Tables S4 and S5 of the SI. Notably, all tested wave function approaches, except for CCSD, yield a negative MSE, indicating an overestimation of the 
magnitude of the (negative) STGs. This trend is particularly pronounced for CIS(D), and the four ``spin-scaled'' second-order approaches. In contrast, CCSD tends to provide smaller STG estimates, with no inversion observed for 
both {\bfseries{7}} and {\bfseries{8}}, for which our TBEs are slightly negative. As one can see by comparing the MSEs and MAEs, these trends are systematic. All wave function approaches, except for CIS(D), provide acceptable SDEs, 
correctly predicting the chemical trends in the series, which is a satisfying outcome. Among these methods, the most refined scheme, CCSD, produces the smallest SDE, while, as expected, \cite{Ver21} the two spin-scaled CC2 models 
also provide more consistent values than the standard CC2 approach. In contrast, at the ADC(2) level, only the SCS approach improves the SDE.

Taking into consideration the usual trade-off between absolute and relative accuracies, as well as the computational cost, ADC(2) stands out as the most suitable cost-effective option for assessing the STG in similar systems. It is also noteworthy that 
ADC(2) delivers satisfactory absolute singlet and triplet energies (see Table S4 and S5 in the SI). Shifting our focus to the TD-DFT results, it is important to highlight that PBE0-2 fails to provide a reasonable triplet energy
for the (challenging) compound {\bfseries{6}}, which very badly affects its otherwise satisfactory statistics. All three QIDH-derived functionals tend to magnify the negative character of the STG and yield relatively large SDEs, equal to or
larger than that of CIS(D). Among the double-hybrid functionals tested, SCS-PBE-QIDH proves to be the most reliable, though it offers less consistent results when compared to ADC(2).
 
\begin{table}[htp]
\caption{\footnotesize MSE, MAE, and SDE (in eV) determined for the STG, considering our TBE of Table S1 as reference. }
\vspace{-0.2 cm}
\label{Table-4}
\footnotesize
\begin{tabular}{p{3.8cm}p{1.4cm}p{1.4cm}p{1.4cm}}
\hline
Method						& MSE 	&MAE	&SDE	\\
\hline		
CIS(D)						&$-0.253$	&0.253	&0.083	\\
ADC(2)						&$-0.033$	&0.033	&0.035	\\
SOS-ADC(2)				&$-0.193$&0.193	&0.089	\\
SCS-ADC(2)				&$-0.115$&0.115	&0.019	\\
CC2							&$-0.027$	&0.027	&0.041	\\
SOS-CC2						&$-0.165$	&0.165	&0.031	\\
SCS-CC2						&$-0.111$	&0.111	&0.021	\\
CCSD						&$+0.081$&0.081	&0.014	\\
PBE0-2$^a$					&$-0.066$&0.114	&0.235	\\
SOS-PBE-QIDH				&$-0.073$	&0.075	&0.084	\\
SCS-PBE-QIDH				&$-0.033$	&0.055	&0.085	\\
SOS-RSX-QIDH				&$-0.084$	&0.105	&0.121	\\
\hline
\end{tabular}
\vspace{-0.3 cm}
\begin{scriptsize}
\begin{flushleft}
{$^a$For PBE0-2, {\bfseries{6}} is a clear outlier due to a strong orbital mixing in the triplet state (see the SI). Removing it yields MSE, MAE, and SDE of $0.006$, $0.046$, and $0.056$ eV, respectively. Note, however, that
removing this challenging compound would improve the statistics of all other double-hybrid functionals.}
\end{flushleft}
\end{scriptsize}
\end{table}

Finally, we also consider cyclborane, the equivalent of cyclazine {\bfseries{2}} but with a central boron atom instead of a nitrogen atom. This compound has been previously studied by Dreuw and Hoffmann. \cite{Dre23} As detailed in Section S3 of the SI, the $D_{3h}$ structure presents an inverted STG. Yet this GS geometry is not a genuine minimum (one imaginary frequency at the MP2 level) and there exists a $C_{3h}$ GS structure, approximately 1 kcal.mol$^{-1}$ more stable than the $D_{3h}$ conformer
at the CCSD(T)/cc-pVTZ level. This $C_{3h}$ structure exhibits a clear bond length alternation and conforms to Hund's rule (Table S6). Consequently, this structure was not included in the benchmark discussed above.

In short, this contribution conclusively demonstrates that compounds {\bfseries{1}}--{\bfseries{6}}, {\bfseries{9}}, and  {\bfseries{10}} do indeed exhibit inverted vertical STGs. While {\bfseries{7}} and  {\bfseries{8}} also deviate
from Hund's rule according to our TBEs, the amplitudes of their negative STG are likely too small to be definitive.  For both the lowest triplet and singlet states of all these compounds, we performed CC3/CCSDT 
calculations with various diffuse-containing basis sets, as no indications of multireference character or a double-excitation nature were detected. For heptazine, neither the relaxation of the geometry in the excited states nor the inclusion of 
vibrational corrections alters the conclusion reached in the vertical approximation. In other words, the inverted STG persists when considering 0-0 energies. 

When assessing the performance of lower-order methods suitable for computing the properties of substituted or extended triangulenes, using our TBE values, we found that ADC(2) likely strikes the best balance between accuracy, consistency, and cost among the wave function methods. 
While less consistent than ADC(2), the SCS-PBE-QIDH double-hybrid functional shows promise for TD-DFT calculations. 
We hope that this Letter can serve as the starting point for the accurate design of systems displaying inverted STGs.

\section*{Computational details}

\subsection*{General}

All calculations rely on the frozen-core (FC) approximation and, except when noted, the default convergence thresholds and algorithms of a given code. In the text,
we dropped the EOM/LR prefix for the CC methods since the two formalisms provide the same transition energies.

\subsection*{Geometries}

All GS structures were first optimized at the MP2/6-311G(d,p) level with \textsc{gaussian} 16.A.03, \cite{Gaussian16} using a $Z$-matrix enforcing the
highest possible point group symmetry. Tight convergence thresholds were applied during these optimizations. Analytical frequency calculations were next performed at the 
same level of theory, confirming that the structures are true minima. These stable MP2(FC)/6-311G(d,p) geometries were considered as starting point for CCSD(T)/cc-pVTZ 
analytical optimizations that were performed with \textsc{cfour}.\cite{cfour,Mat20} 

For the 0-0 calculations on heptazine, the GS geometry was reoptimized at the CCSD/cc-pVDZ level of theory, and vibrational frequencies were determined as well
at this level using \textsc{gaussian}.\cite{Gaussian16}  The structure and vibrational frequencies of the lowest triplet were obtained with the same code at the UCCSD/cc-pVDZ level
whereas we applied the corresponding EOM-CCSD/cc-pVDZ level for the lowest singlet ES. Symmetry was lowered when imaginary frequencies were obtained, 
and the process was restarted until a true minimum was reached.

\subsection*{Transition energies}

We used a variety of quantum chemistry software to determine vertical transition energies, using three gaussian basis sets containing both polarization and diffuse functions, namely 6-31+G(d),
\emph{aug}-cc-pVDZ, and \emph{aug}-cc-pVTZ. This allowed us to obtain TBEs, following the formulas given in the footnotes of Tables \ref{Table-1} and \ref{Table-2}, which assume that
basis set effects are transferable at the CC levels, an approximation that we have extensively assessed and validated in previous work. \cite{Loo22}
CCSD calculations have been performed with \textsc{dalton} \cite{dalton} and \textsc{gaussian}. \cite{Gaussian16}  CC3  \cite{Chr95b}  
calculations have been performed with \textsc{dalton} \cite{dalton} as well as \textsc{cfour}, \cite{cfour,Mat20} the latter allowing calculations to singlet ES only. CCSDT  \cite{Kow01} 
calculations were also carried out with \textsc{cfour} and were only possible with the most compact basis set. The same code was used for the CCSD(T)(a)$^\star$ \cite{Mat16}
and single-point UCCSD(T) calculations with the triple-$\zeta$ basis set.  In \textsc{cfour}, we relied on the QC-SCF algorithm \cite{Not21} with the correct occupation number set, and a SCF convergence threshold 
of $10^{-9}$ or $10^{-10}$ au.  All CIS(D), \cite{Hea94} ADC(2), \cite{Dre15} and CC2 \cite{Chr95} calculations have been performed with \textsc{turbomole} \cite{Bal20,Turbomole} using the 
\emph{aug}-cc-pVTZ basis set and applying the RI approach \cite{Hat00} with the corresponding auxiliary basis.  In these calculations, we enforced the default \textsc{turbomole} scaling parameters
for both SOS-ADC(2), SCS-ADC(2), SOS-CC2 and SCS-CC2. The TD-DFT calculations based on double hybrids were all performed with \textsc{orca}, \cite{Nee20} selecting  PBE0-2, \cite{Cha12c} SOS-PBE-QIDH, 
\cite{Cas21b} SCS-PBE-QIDH, \cite{Cas21b} and SOS-RSX-QIDH,  \cite{Cas21b} and also using the FC approach. In \textsc{orca}, we set the \texttt{tightSCF} and \texttt{grid3} options, with the   \emph{aug}-cc-pVTZ 
and the automatically generated auxiliary basis sets. Note that the Tamm-Dancoff approximation was not enforced in the TD-DFT calculations.

\section*{Acknowledgments}
The authors are thankful for the generous allocations of time by the CCIPL/GliCID computational center installed in Nantes.
PFL thanks the European Research Council (ERC) under the European Union's Horizon 2020 research and innovation programme (Grant agreement No.~863481) for funding.

\section*{Supporting Information Available}
Full benchmark results; Molecular orbital plots; Cyclborane results; Cartesian coordinates. 

\bibliography{biblio-new}
\end{document}